\documentclass[aps,prl,twocolumn,twoside,superscriptaddress,notitlepage]{revtex4-2}

\usepackage{tikz}
\usetikzlibrary{calc}
\usepackage{mathtools}
\usepackage{graphicx}
\usepackage[dvipsnames]{xcolor}
\usepackage{ulem, enumitem}
\usepackage{bbm, bm}
\usepackage{amsmath, amssymb, amsfonts,dsfont}
\usepackage{natbib}
\usepackage[colorlinks=true,citecolor=blue,linkcolor=blue,urlcolor=blue]{hyperref}

\newtheorem{theorem}{Theorem}

\newenvironment{proof-of}[1]{\medskip\noindent\textbf{Proof of {#1}.}}{\hfill$\blacksquare$\medskip}

\newcommand{\ket}[1]{\left\vert#1\right\rangle}

\newcommand{\T}{\mathrm{T}}

\begin{document}

\title{Bounding Kirkwood-Dirac negativity of Gaussian processes}

\author{Luca Bianchi}
\affiliation{Department of Physics and Astronomy, University of Florence, 50019, Firenze, Italy}
\email{luca.bianchi@unifi.it}

\author{Carlo Marconi}
\affiliation{Istituto Nazionale di Ottica del Consiglio Nazionale delle Ricerche (CNR-INO), 50125 Firenze, Italy}

\author{Riccardo Cioli}
\affiliation{
Dipartimento di Fisica e Astronomia, Università di Bologna and INFN, Sezione di Bologna, via Irnerio 46, I-40126 Bologna, Italy
}

\author{Jan Sperling}
\affiliation{Paderborn University, Institute for Photonic Quantum Systems (PhoQS), Theoretical Quantum Science, Warburger Stra\ss{}e 100, 33098 Paderborn, Germany}

\begin{abstract}
    The Kirkwood-Dirac quasiprobability provides an operational representation of a quantum state, whose negativity serves as a measure of nonclassicality.
    Despite its fundamental importance, the extremal values of the Kirkwood-Dirac negativity are still unknown in the general case.
    We investigate the Kirkwood-Dirac quasiprobability of an arbitrary quantum state under Gaussian processes.
    In this setting, we derive an upper bound on the negativity for any number of modes and measurements.
    For a single mode and two measurements, we show that the eigenstates of the quadrature operators saturate this upper bound, while a nontrivial minimum is reached by pure Gaussian states.
    As a consequence, our results indicate that Gaussian states are sufficient to achieve extreme values of nonclassicality.
\end{abstract}

\date{\today}
\maketitle

\paragraph{Introduction.---}
    Quantum information processing exploits uniquely quantum features to encode, to manipulate and to retrieve information in ways that may surpass classical capabilities.
    Among the different quantum information platforms, continuous-variable (CV) systems encode information in observables with continuous spectra \cite{brauvanloock}.
    Such observables arise naturally in a variety of physical systems, including the quadratures of the electromagnetic field \cite{mandel1995optical}, collective spin variables of atomic ensembles \cite{cerf2007quantum}, and the motional degrees of freedom of mechanical oscillators \cite{barzanjeh2022optomechanics}.
    
    Within the CV framework, quantum states and operators admit a phase-space representation in terms of functions on a symplectic vector space.
    The intrinsic quantum features of such representations may be captured by looking at quasiprobability distributions \cite{ferrie2011quasi}, which generalize classical probability distributions by relaxing Kolmogorov's axioms \cite{Kolmogorov1960FoundationsOT}. 
    As a result, quasiprobabilities may assume negative or even complex values, providing signatures of nonclassicality.
    This property has made them a powerful framework for the description of genuinely quantum properties, such as entanglement \cite{sperling2009representation}, contextuality \cite{spekkens2008negativity}, coherence \cite{sperling2018quasiprobability}, and non-Gaussianity \cite{albarelli2018resource}, representing a viable tool for the certification of these quantum resources.
    
    %Among them, non-Gaussianity plays a particularly prominent role.
    %Although Gaussian states and operations constitute the basic building blocks of CV quantum information, owing to their efficient generation and manipulation \cite{wang2007quantum}, quantum non-Gaussianity has been identified as a necessary ingredient for achieving quantum advantage \cite{bartlett2002efficient}.
    %Its characterization can be approached through quasiprobability representations in several ways \cite{genoni2010quantifying}, for example via non-Gaussian monotones \cite{genoni2007measure}, higher-order cumulants \cite{bednorz2011fourth}, or the stellar rank \cite{chabaud2020stellar}.
    
    Beyond the well-known family of $s$-parametrized quasiprobability distributions \cite{vogel2006quantum}, the Kirkwood–Dirac (KD) quasiprobability \cite{kirkwood1933quantum, dirac1945analogy} has recently attracted considerable attention.
    It has been shown to play a significant role in the study of coherence \cite{budiyono2023quantifying, budiyono2024quantum}, entanglement \cite{budiyono2025quantum}, contextuality \cite{de2023relating, schmid2024kirkwood}, metrology \cite{arvidsson2020quantum}, weak values \cite{wagner2024quantum}, and quantum thermodynamics \cite{gherardini2024quasiprobabilities}
    (see \cite{arvidsson2024properties} for a comprehensive review).
    Unlike $s$-parametrized quasiprobabilities, where the phase-space structure is fixed and depends on the Heisenberg--Weyl Lie algebra, in the case of KD distributions, its associated space can be defined operationally through sequences of generalized weak measurements of non-commuting observables.
    As a consequence, this framework has found application in the implementation of direct measurements of quantum states \cite{lundeen2011direct, lundeen2012procedure,bamber2014observing, thekkadath2016direct,lostaglio2023kirkwood}.
    Since the choice of measurements is largely arbitrary, KD formalism has proven to be a particularly versatile tool for probing generic quantum features.
    Previous works \cite{lundeen2012procedure,bamber2014observing} established a connection between the KD distribution of quadature operators and the Agarwal--Wolf quasiprobabilities \cite{agarwal1970calculusLE1, agarwal1970calculusLE2}, thus taking the first step towards the description of KD quasiprobability in CV quantum systems.
    
    Nevertheless, a complete characterization of the KD distribution for Gaussian processes is still lacking. 
    Given the central role of Gaussian states and operations in a variety of quantum information protocols, further investigation of this connection is highly valuable. 
    In particular, it would be interesting to determine the range of KD negativity attainable within the Gaussian regime and to identify the states that maximize this resource.
    This is in analogy to the role of Bell states in achieving maximal Bell nonlocality \cite{braunstein1992maximal}, finding minimal uncertainty states \cite{bergou1991, pegg2005minimum}, etc. 
    
    In this work, we fill this gap by providing the KD distribution for an arbitrary Gaussian process and quantifying its negativity.
    Specifically, we prove that, similarly to the case of the Wigner function, the KD negativity can be upper-bounded for any number of modes and measurements.
    This bound corresponds to a simple function of the covariance matrices of the Gaussian measurements considered.
    For a two-measurement, single-mode Gaussian process, such a bound is attained by properly tailored quadrature eigenstates.
    A global minimum can also be found for pure states and, similarly to the supremum, its magnitude is governed by the determinant of the sum of covariances.
    To find such extremal points, we harness the correspondence between single-mode covariance matrices and events on a $(2+1)$-Minkowski spacetime, reducing the problem to the kinematics of a particle in the center of mass reference frame.
    Finally, we consider two families of non-Gaussian states and investigate their capability in saturating the upper bound of the KD negativity.
    Our findings indicate that such states never attain the maximum, confirming that Gaussian states are sufficient to achieve extreme values of the KD negativity.
    
\paragraph{Kirkwood-Dirac for Gaussian processes.---}
    \label{sec:gaussian}
        
        Consider an $M$-mode quantum state $\hat{\rho}$ undergoing $N$ sequential measurements 
        $
        \{\hat{\Pi}_i\}_{i=1}^N
        $,
        where each $\hat{\Pi}_i$ is an element of a set of positive operator-valued measures (POVMs).
        Between two consecutive measurements, the time evolution acts by means of a set of unitary operators
        $
        \{\hat{U}_{G_i}\}_{i=1}^{N-1}
        $, where $G$ refers to the quadratic Hamiltonians driving the system.
        According to the theory of quasiprobabilities \cite{ferrie2011quasi}, the set of operators 
        \begin{equation}
            \label{def:frame}
                    \hat F(r)
                    =
                    \hat{U}^{\dagger}_{G_1}
                                \dots
                    \hat{U}^{\dagger}_{G_{N-1}}
                    \hat{\Pi}_N\hat{U}_{G_{N-1}}\hat{\Pi}_{N-1}
                                \dots \hat{U}_{G_1}\hat{\Pi}_1~,
        \end{equation}
        for different values of $r$, defines a frame for the Hilbert space of states. Here $r = (r_1,\dots,r_N)$ is a vector whose components correspond to the measurement outcomes.
        This frame completely characterizes a given Gaussian process and can be used to define the KD distribution as 
        \begin{equation}
            \label{def:KD}
                Q(r) = 
                    \mathrm{Tr}
                    \left[
                        \hat F(r)
                        \hat\rho
                    \right].
        \end{equation}
        Since the operator $\hat F(r)$ is, in general, neither positive-semidefinite nor hermitian, the function $Q(r)$ can assume negative and complex values, which are a sign of nonclassicality \cite{arvidsson2024properties}.
        To quantify KD nonclassicality, one can resort to Eq.~\eqref{def:KD} the total negativity given b
        \begin{equation}
            \label{eq:def_negativity}
                \mathcal{N} 
                =
                    \int_{\Gamma^N} dr
                            \left|
                                Q(r)
                            \right|~.
        \end{equation}
        In particular, $\mathcal{N}=1$ if $Q$ is a classical probability distribution, while $\mathcal{N}>1$ signals the nonclassicality of the associated quantum state.

        An $M$-mode phase space can be introduced as the pair 
        $\Gamma = (\mathbb{R}^{2M},\Omega)$, where $\Omega = \bigoplus_{j=1}^M\left(\begin{smallmatrix}0 & 1 \\ -1 & 0\end{smallmatrix}\right)$ is the symplectic form. 
        Hence, the configuration space is just its $N$-fold version, i.e.,
        $
        \Gamma^{N} = [(\mathbb{R}^{2M})^{\times N}, \Omega_N]
        $, 
        with $\Omega_N = \bigoplus_{i=1}^N \Omega$.
        In the phase space formalism, operators can be described by functions of canonical variables $q = (x_1,p_1,\dots,x_M,p_M)$, where $(x_{i}, p_{i})$ correspond to a pair of conjugate variables for the $i$-th mode.
        For Gaussian processes, the dynamics is generated by Hamiltonians which are at most quadratic in the bosonic field operators $\left\{\hat{a}_j\right\}_{j=1}^M$, with $\hat{a}_j = (\hat{x}_j+i\hat{p}_j)/\sqrt2$.
        Such Hamiltonians generate unitaries that act as symplectic matrices on phase space vectors.
        For bosonic operators, one can adopt the so-called Weyl representation to map operators into phase space functions. 
        This representation relies on displacement operators of the form $\hat{D}_q = \exp{\left[ iq^{\T}\Omega\hat{q}\right]}$,
        where
        $\hat{q} = (\hat{x}_1,\hat{p}_1,\dots,\hat{x}_M,\hat{p}_M)$ denotes the vector of quadrature operators. 
        These operators form a complete orthogonal basis for Hilbert space of bounded operators with respect to the Hilbert-Schmidt inner product. 
        Therefore, any operator $\hat{A}$ on this space can be expressed in terms of a characteristic function, $\chi_{A}(q) = \mathrm{Tr}\left[\hat{A}\hat{D}_{-q}\right]$.
        %\begin{equation}
         %   \hat{A} = \int_{\Gamma} \frac{dq}{(2\pi)^M} \chi_{A}(q) \hat{D}_q,
        %\end{equation}
        %where $\int_{\Gamma}dq$ denotes the integration over the entire phase space and the characteristic function $\chi_A$ is defined as
        %\begin{equation}
         %   \chi_{A}(q) = \mathrm{Tr}\left[\hat{A}\hat{D}_{-q}\right].
        %\end{equation}
        %\luca{allowing the computation of arbitrary expectation values.}
        Within this setting, Gaussian operators are those operators with a Gaussian characteristic function, i.e.,
        \begin{equation}
            \label{def: gausschar}
            \chi_{A_{i}}(q_i) = 
                \exp{
                    \left[
                        -\frac{1}{4}q_i^{\T}\Omega^{\T}\sigma_i \Omega q_i + i q_i^{\T} \Omega^{\T} r_i
                    \right]
                }, 
        \end{equation}
        where $r_i$ and $\sigma_i$ denote, respectively, the mean value and the covariance matrix associated to the operator $\hat{A}_i$.
        
        In the case of Eq.~\eqref{def:frame}, the operators $\{\hat{\Pi}_i\}_{i=1}^N$ describe projectors onto Gaussian states characterized by covariance matrices $\sigma_i$ and measurement outcomes $r_i$, while $\{\hat{U}_{G_i}\}_{i=1}^{N-1}$ are Gaussian unitaries corresponding to symplectic rotations $G_i$.
        For later purposes, it is also useful to introduce $\mathrm{G}_j = G_{j-1} \dots G_0$.
        For consistency, we assume $G_1 = G_0 = \mathrm{id}_{2M}$, where $\mathrm{id}_{2M}$ is the $2M \times 2M$ identity matrix in phase space.
        The frame operator $\hat{F}(r)$ defines an $N$-measurements multi-mode Gaussian process for which we prove the following result:
        \begin{theorem}
            \label{prop:global_up}
            The KD negativity of an arbitrary quantum state undergoing an $M$-mode Gaussian process with $N$ measurements, satisfies 
            \begin{equation}
                \label{eq:upper_bound}
                \mathcal{N}
                \leq 
                (2\pi)^{MN}
                \sqrt[4]{
                    \prod_{i=1}^{N-1}
                    \det[ 
                    \sigma^{(\mathrm{G}_i)}+\sigma^{(\mathrm{G}_{i+1})}    
                    ]
                }~,
            \end{equation}
            with  
            $
            \sigma^{(\mathrm{G}_j)} 
            = 
            (\mathrm{G}\mathrm{G}^{\T})^{-\T}_j\sigma_j(\mathrm{G}\mathrm{G}^{\T})_j
            $.
        \end{theorem}
        We refer to Section 1 of the Supplemental Material (SM) \cite{SM} for the proof of Theorem \ref{prop:global_up}.
        
        Interestingly, the upper bound in Eq.(\ref{eq:upper_bound}) depends only on the covariance matrices of each pair of consecutive measurements. Note that this reflects the structure of the sequential measurements in Eq.~\eqref{def:frame}.
        If one additionally assumes an input Gaussian state with covariance matrix $\sigma$, the KD distribution takes the form
        \begin{equation}
            \label{eq:KDgauss}
                Q(r) = \frac{2^{MN}}{\sqrt{\det[\Lambda]}}
                    e^{
                            - r^{\T}
                            \Lambda^{-1}
                            r 
                        }~,
        \end{equation}
        where the explicit form of Eq.(\ref{eq:KDgauss}) has been derived in Section 2 of the SM \cite{SM}.
        Here,  $\Lambda$ is a $2NM \times 2NM$ block matrix,
        \begin{equation}
            \label{def:lambda}
                \Lambda = 
                \begin{bmatrix}
                    \sigma +\sigma^{(\mathrm{G}_1)}& \sigma -i\Omega &  \dots & \sigma -i\Omega\\
                    \sigma +i\Omega& \sigma +\sigma^{(\mathrm{G}_2)} & \dots & \sigma -i\Omega\\
                    \vdots & \vdots & \vdots & \\
                    \sigma +i\Omega& \sigma +i\Omega& \dots &\sigma +\sigma^{(\mathrm{G}_N)}\\
                \end{bmatrix},
        \end{equation}
        while $r = (r_1,\dots,r_N)$ is an $N$-dimensional vector of $2M$-dimensional variables in the full phase space $\Gamma^N$.
        For the sake of simplicity and without loss of generality, the displacement vector for the input state has been set to zero. Note that
        this does not affect the value of the negativity Eq.~\eqref{eq:def_negativity}, which is invariant under translations in phase space.
        Moreover, we observe that the negativity of the KD distribution is determined by the off-diagonal blocks of the complex-symmetric matrix $\Lambda$.
        %Such blocks mix the variables $r_i$ belonging to different measurements $\hat{\Pi}_i$ and are responsible for the incompatibility between Gaussian POVMs. 
        Taking $N=1$, one readily obtains a positive probability distribution, e.g., an unnormalized Husimi function for $\hat\Pi_1 = \hat{D}_1$.
        For $N=2$, instead, one can recover the Agarwal-Wolf quasiprobability in the $xp$ (standard) ordering \cite{agarwal1970calculusLE1,agarwal1970calculusLE2}, by taking the limits $\sigma_x = \lim_{\zeta_1\rightarrow \infty} \sigma_1 $ and $\sigma_p = \lim_{\zeta_2\rightarrow 0} \sigma_2$, for each mode with squeezing $\zeta$. Note that such limits correspond to the case of homodyne detection.

        In the general case of a Gaussian KD distribution in Eq. \eqref{eq:KDgauss}, we obtain a closed formula for the negativity, i.e.,
        \begin{equation}
            \label{eq:KD_final}
            \mathcal{N}_g 
            = 
            (2\pi)^{MN}
            \sqrt{
                \frac{|\det[\Lambda]|}{\det[\mathrm{Re}[\Lambda]]}
            }.
        \end{equation}  
        For the sake of simplicity, in what follows, we will restrict to the case $N=2$. This choice is nevertheless experimentally relevant, since two-measurement scenarios are sufficient to account for the vast majority of weak measurements implemented in laboratories \cite{lundeen2011direct,lundeen2012procedure,arvidsson2024properties}. 
        We prove in Section 3 of the SM \cite{SM} that the KD negativity is a Loewner non-increasing monotone for Gaussian states. As a consequence, the optimization can be restricted to the set of pure states with covariance matrices saturating the minimum uncertainty condition.  
        Within the above setting, we prove the following Theorem \ref{prop:global_up}:
        
        \begin{theorem}
            \label{th:neg_max}
    For any Gaussian process with $M=1$ and $N=2$, the global maximum of the KD negativity of Theorem \ref{prop:global_up} is attained by any Gaussian state in the limit of infinite squeezing, for a phase $\phi^{\mathrm{max}}(\sigma_1,\sigma_2)$.
    Moreover, within the set of pure Gaussian input states, the minimum KD negativity is given by
    \begin{equation}
    \mathcal{N}_g^{\mathrm{min}}     =
    (2\pi)^2 
    \sqrt{1 -\left(\sqrt{\det[\sigma_1+\sigma_2]} + 1\right)^{-2}}~,
    \end{equation}
    which is achieved for a pair of squeezing and phase parameters 
    $(\zeta^{\mathrm{min}}(\sigma_1,\sigma_2),\phi^{\mathrm{min}}(\sigma_1,\sigma_2))$.
        \end{theorem}
             We refer the reader to Section 4 of the SM \cite{SM} for the proof of Theorem 2 and the explicit values of $\phi^{\mathrm{max}}$,$ \phi^{\mathrm{min}}$ and $\zeta^{\mathrm{min}}$.
             Note that the proof heavily relies on mapping the problem onto the two-body kinematics in the center-of-mass frame of a $(2+1)$-dimensional Minkowski spacetime.

\paragraph{Application: KD distributions for non-Gaussian CV states.---}
    \label{sec:KD_non_gaussian}
    
        To assess our findings, we compute the maximum value of the KD negativity for several classes of quantum states and compare it with the upper bound in Theorem \ref{prop:global_up}.
        We choose pure Gaussian measurements such that, according to their symplectic decomposition, the squeezing parameters are $\zeta_1 = \zeta_2 = 4$, corresponding to a squeezing of $3 \, \mathrm{dB}$, while their angles are $\phi_1 = 0$ and $\phi_2 = \pi/4$. 
        To display negativities and to reveal the non-trivial structure of the KD distributions, amplitudes and phases are presented for phase space slices of fixed $(x_i,p_i)$.
        %Note that marginals of KD distributions are proper probabilities, not showing quantum features \cite{arvidsson2024properties}.
        
        Consider the squeezed Fock states, described by
        \begin{equation}                    
            \ket{S,n} = \hat{U}_S\ket{n},
        \end{equation}
        where $\hat{U}_S$ is a Gaussian unitary.
        Their characteristic function can be written in terms of the derivatives of a Gaussian generating function \cite{sperling2012true,engelkemeier2020quantum,bianchi2025unified}, i.e.,
        \begin{equation}
        \label{char_func}
            \chi^{(n)}(q) 
            = 
            \left.
                \frac{
                    \partial^{n}_{z}
                    }{n!}
                \frac{
            \exp{
                \left[
                    -\frac{1}{4}
                    q^{\T}
                    \Omega^{\T}
                    \eta(z)
                    \Omega
                    q
                \right]
                }
            }{\prod_{j=1}^M(1-z_j)}
            \right|_{z = 0},
        \end{equation}
        where 
        $
            \eta(z) 
            = 
            S^{-1}
            \left[\bigoplus_{j=1}^M
                \frac{1+z_j}{1-z_j}
                \mathrm{id}_2
            \right]
            S^{-\T}
        $,
        $|z_j| < 1 $ for all $ j\in\{ 1,\ldots,M\}$ and a multi-index notation is introduced for $n! = n_1!\dots n_M!$ and $\partial_{z}^n = \partial_{z_1}^{n_1}\dots\partial_{z_M}^{n_M}$.
        %Note that $\eta(z)$ is not the covariance matrix of a pure Gaussian quantum state anymore, matching the well-known fact that Fock states are not minimal uncertainty states and can be obtained by a Gaussian characteristic function, being a form of generating function .
        
        By combining Eq.~\eqref{char_func} with the Gaussian process analysis derived in Section 2 of the SM \cite{SM}, the KD distribution reads
        \begin{equation}
            \label{eq:KD_photons}
                Q^{(n)}(r) 
                =
                \left.\frac{\partial^{n}_{z}}{n!}
                \frac{2^{MN}
                \exp{
                            \left[
                                - r^{\T}\Lambda(\zeta, z)^{-1}r 
                                \right]
                            }
                }{
                    \sqrt{
                    \det{
                        \Lambda(\zeta,z)}}
                \prod_{j=1}^M(1-z_j)
                } 
                \right|_{z = 0},
        \end{equation}
        where we have replaced $\Lambda \to \Lambda(z)$ for $\sigma \to \eta(z)$.
        
\begin{figure}
            \includegraphics[width=\linewidth, trim={4.2cm 1.2cm 2.3cm 4.5cm},clip]{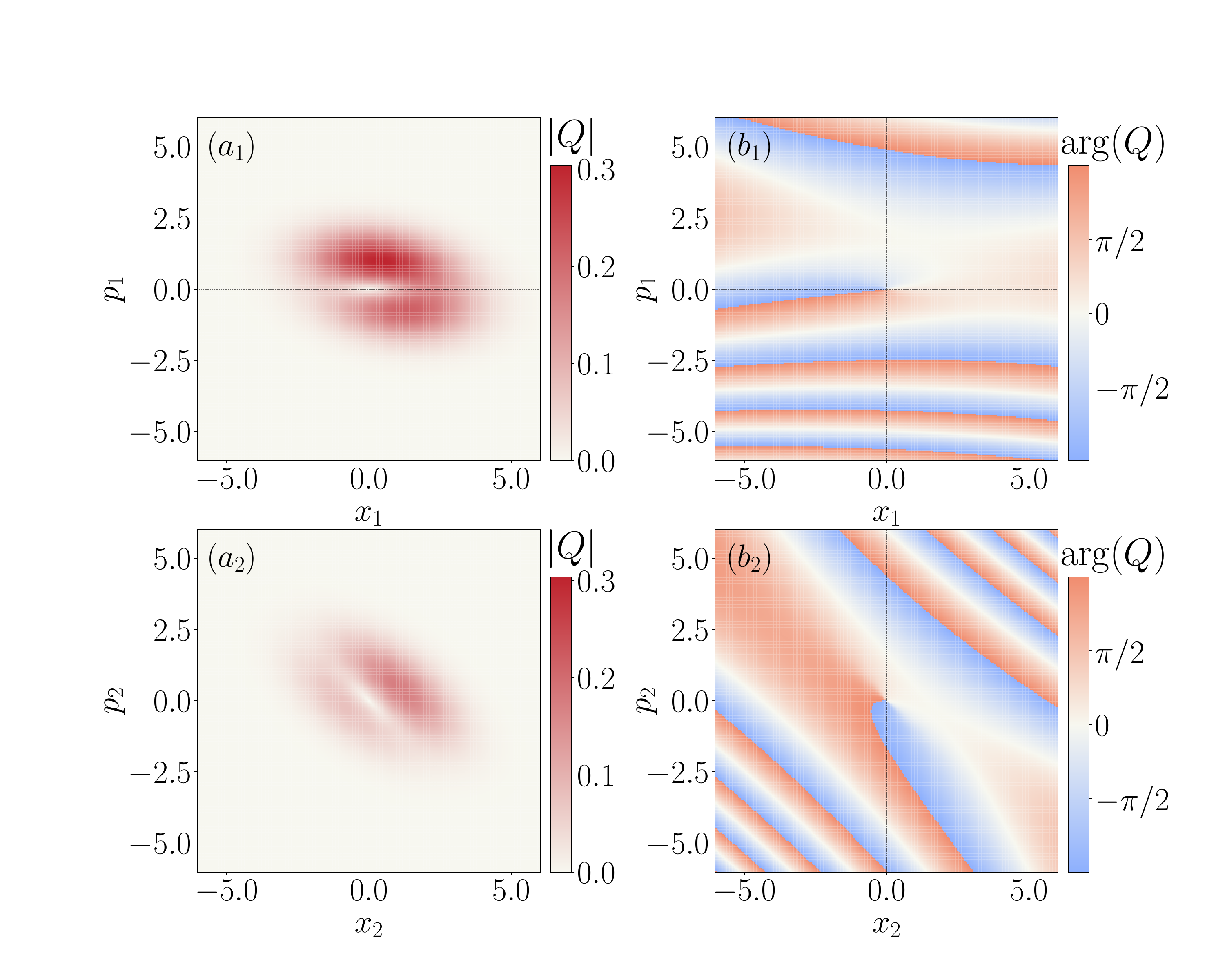}
            \caption{
                $Q(x_1,p_1,x_2,p_2)$ for a single-Fock state.
                The different subplots show the amplitude (left) and the phase (right) of the KD distribution for the slices $(x_2,p_2)=(1.25, 0.25)$ (top panels) and $(x_1, p_1) = (1.25, 0.25)$ (bottom panels). 
            }\label{fig:fock_plot}
        \end{figure}

        As an example, consider Fock states, corresponding to $S = \mathrm{id}_2$. 
        The KD distribution for a single-photon is shown in Fig. \ref{fig:fock_plot} as a 2-dimensional slice of the 4-dimensional phase space.
        From the value of $\arg(Q)$, it is evident that the KD distribution is highly nonclassical.
        %The negativity shows to saturate the bound of Proposition \ref{prop:global_up} in the limit of large Fock number $n$.
        
%%%%%%%%%%%%%%%%%%%%%%%                
                %The KL distance of non-Gaussianity and the HS-based non-Gaussianity criterion are shown in fig. .
                %Here, the closest Gaussian state in the Hilbert-Schmidt distance has moments with moments
                %\begin{equation}
                 %   \begin{split}
                  %      &\bar q = 0\\
                   %     &\sigma 
                    %    = 
                        %\frac{\partial^%{n}_{z}}
                        %{n!}
                        %\left[
                         %   \frac{1+z}{(1-z)^2}
                         %   SS^{\T}
                        %\right]_{z=0}
                    %\end{split}
                %\end{equation}
                %The comparison between %the two quantities is %shown in fig. \ref{fig:nongauss_fock}.
                %Both criteria show a monotonically increasing behavior with increasing Fock number $n$, highlighting  the deep non-Gaussianity of Fock states of increasing size.
                %\begin{figure}
                    %\centering
                    %\includegraphics[width=\linewidth]{KL_vs_delta_fock.png}
                    %\caption{
                    %Kullback-Leibler divergence (red, solid line) compared with Hilbert-Schmidt distance (blue, dashed line) as functions of Fock number $n$.
                   % Both the curves show a monotonically increasing profile with $n$.
                   % }
                    %\label{fig:nongauss_fock}
               % \end{figure}
%%%%%%%%%%%%%%        

\begin{figure}
            \includegraphics[width=\linewidth, trim={4.2cm 1.2cm 2.2cm 4.5cm},clip]{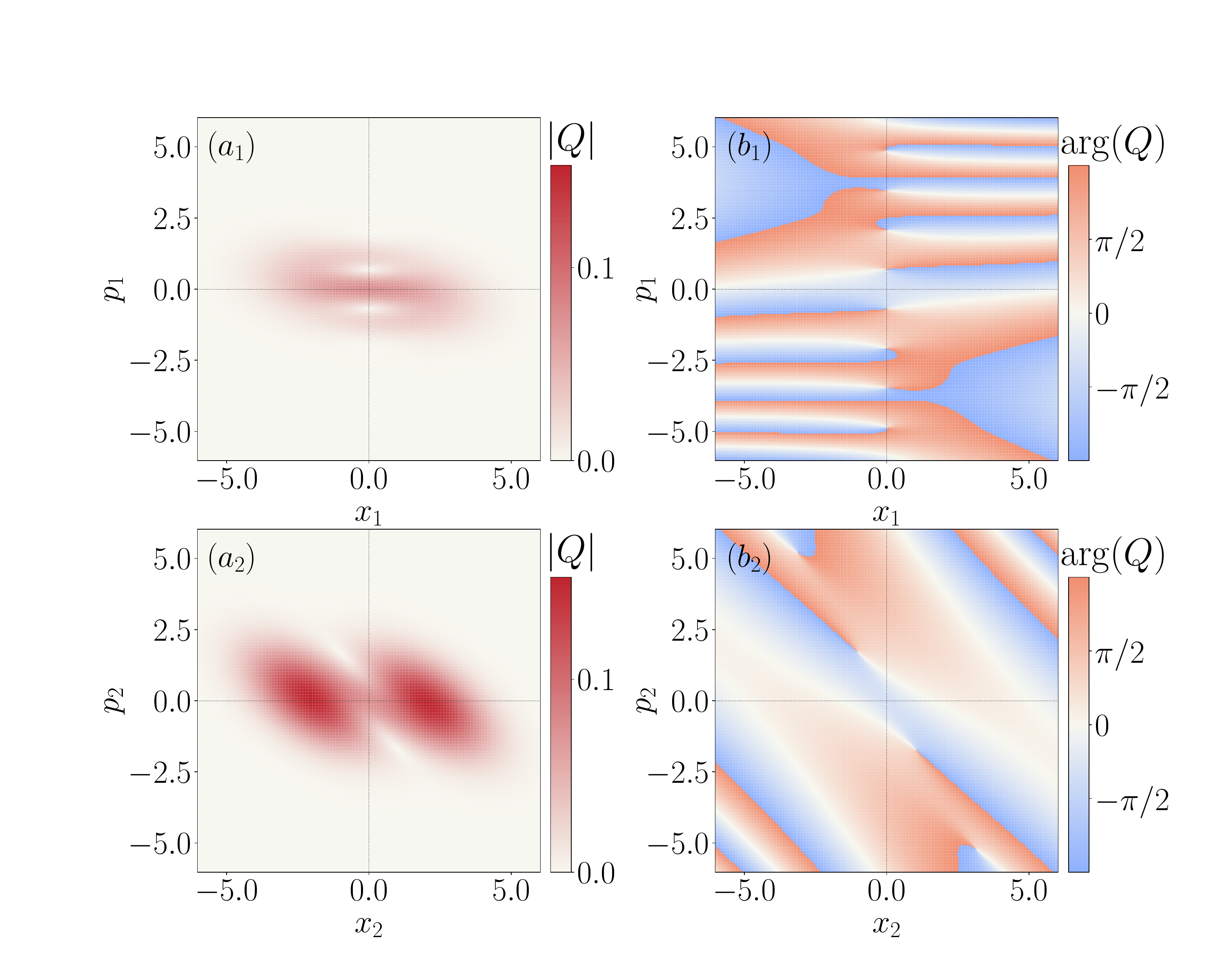}
            \caption{
            $Q(x_1,p_1,x_2,p_2)$ for an even cat state, where $\alpha = 2$. 
            The plots show the amplitude (left) and the phase (right) of the KD distribution for $(x_2,p_2)=(0, 0)$ (top panels) and $(x_1, p_1) = (0, 0)$ (bottom panels). 
           }\label{fig:cat_KD}
        \end{figure}

        The second class of states considered is given by superpositions of coherent states (see Section 5 of SM \cite{SM} for the derivation of the KD distribution for such states).
        In particular, we restrict to the celebrated even and odd single-mode cat states of amplitude $\alpha$, i.e.,
        \begin{equation}
        \label{def:cat}
            \ket{\mathrm{cat}^{(\pm)}} = \frac{\ket{\alpha}\pm\ket{-\alpha}}{\sqrt{2(1\pm e^{-2|\alpha|^2})}},
        \end{equation}
        whose KD distribution reads
        \begin{equation}
            \label{eq:KD_cat
            }
            \begin{split}
                Q^{(\pm)}(r) 
                =& 
                \frac{2^{NM}
                e^{
                    -r^{\T}
                    \Lambda_{\psi}^{-1}
                    r 
                    }
                }{\sqrt{\det[\Lambda_{\psi}]}(1\pm e^{-2|\tilde\alpha|^2})}\\
                &\times\left( 
                    e^{
                - 2\bar\alpha^{\T}
                    \Lambda_{\psi}^{-1}
                    \bar\alpha 
                }
                    \cosh{
                    \left[ 
                        \gamma
                    \right]
                    }
                \right.
                    \\
                    &\left.\pm
                    e^{
                        -2||\tilde\alpha||^{2}
                        + 2\bar\alpha^{\T}
                        \Omega^{\T}
                    \Lambda_{\psi}^{-1}
                        \Omega
                    \bar\alpha 
                    }
                    \cos{
                    \left[
                        \gamma
                    \right]
                    }
                \right),  
            \end{split}
        \end{equation}
        with $\Lambda_{\psi} = \Lambda$ for $\sigma=\mathrm{id_{2M}}$ and $\gamma=2\sqrt{2}
                        \bar\alpha^{\T}
                        \Omega^{\T}
                        \Lambda_{\psi}^{-1}
                        r$. 
        Here, $
                \tilde\alpha_i = (\mathrm{Re}[\alpha_i], \mathrm{Im}[\alpha_i])
                $
                and 
                $\bar\alpha = (\tilde\alpha, \dots, \tilde\alpha)$.
        In Fig. \ref{fig:cat_KD}, the KD quasiprobabilities of an even cat state is displayed as 2-dimensional slices of the 4-dimensional the phase space.
        Again, we observe that $Q\neq|Q|$ from the non-zero phases almost everywhere.
        
\begin{figure}
            \includegraphics[width=\linewidth, trim={1.2cm 1.5cm 1.2cm 1.2cm},clip]{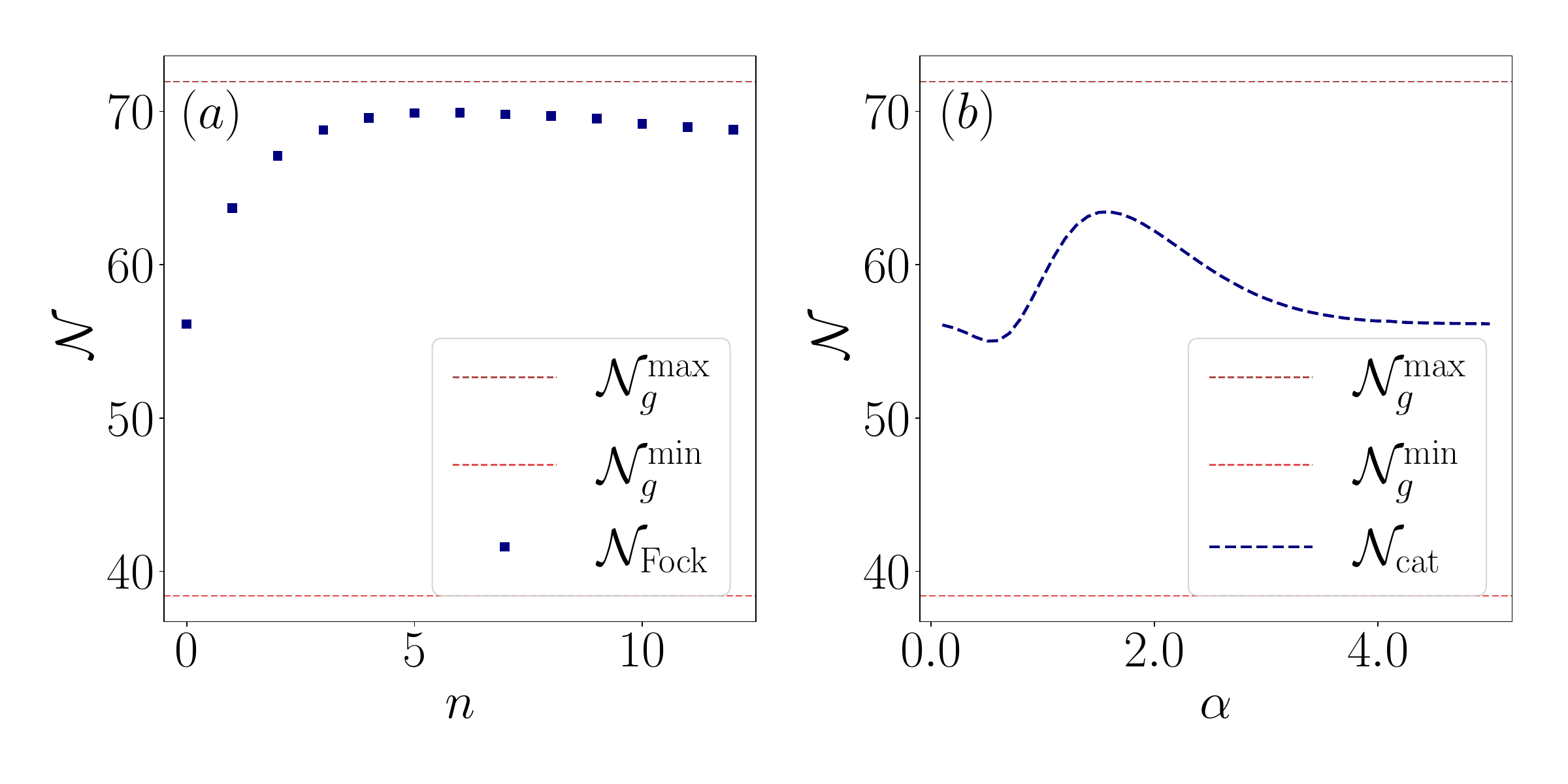}
            \caption{
                Negativity of the KD for non-Gaussian states: (a) Fock states and (b) even cat states.
            }\label{fig:fock_cat_negativity}
        \end{figure}

        The negativity $\mathcal N$ can be plotted as a function of the Fock number $n$ [Fig. \ref{fig:fock_cat_negativity}(a)] and cat's size $\alpha$ [Fig. \ref{fig:fock_cat_negativity}(b)].
        We observe that non-Gaussian states remain within the extremal range given by Theorem \ref{th:neg_max}. 
        In particular, Fock states approach the maximum more closely than cat states.
        In both cases, for small $n$ and $\alpha$, the negativity is close to the vacuum one, which itself is close to the average of the extremal values. 
        As $n$ and $\alpha$ increase, the results show that negativity approaches this average value in the macroscopic limit, thereby indicating that the maximal nonclassicality is located in the intermediate parameter regime.
        As a consequence, our results show that saturation of the upper bound can be achieved with relatively simple states such as the quadrature eigenstates, without resorting to highly non-Gaussian resources. 
        
 %%%%%%%%%%%%%%%%%%%%%%%%%%%%%%%%%%               
                %The proxy Gaussian state for cat states can be built by means of the following moments:
                %\begin{equation}
                 %   \begin{split}
                  %      & \bar q^{(\pm)}_{\mathrm{cat}} = 0\\
                   %     & \sigma^{(\pm)}_{\mathrm{cat}} = 
                    %        \frac{1}{1\pm e^{-2|\tilde\alpha|^2}}
                     %           \left[
                      %              1
                       %             +
                                    %4\tilde\alpha \times \tilde\alpha^{\T}
                                    %\pm
                                    %e^{-2|\tilde\alpha|^2}
                                    %\left(
                                       % 1
                                        %-
                                      %  4\Omega \tilde\alpha 
                                        %\times 
                                        %\tilde\alpha^{\T}
                                        %\Omega^{\T}
                                    %\right)
                                %\right]
                    %\end{split}
                %\end{equation}
                
                %Together with $\hat{\rho}_G^{\mathrm(+)}$, one can plot the distances $D(P^{(+)} \, ||\,|P_G^{(+)})$ and $\delta(\hat{\rho},\hat{\rho}_G)$, as it is shown in \ref{fig:confronto_cat}. 
                %Akin to the Fock case, one can see a general agreement between the two different measures of non-Gaussianity.
                
                %\begin{figure}
                 %   \centering
                    %\includegraphics[width=\linewidth]{KL_div_cat_pos.png}
                    %\caption{
                    %Comparison between a standard non-Gaussianity measure and the Kullback-Leibler divergence-based measure for an even cat state. The two profiles show the same qualitative behavior in the studied region of the cat state size.
                    %}
                    %\label{fig:confronto_cat}
                %\end{figure}
                
%%%%%%%%%%%%%%%%%%%%%%%%%%%%%%%%%%%%%%        
\paragraph{Conclusions and outlooks.---}

        In this work we establish a framework for the KD quasiprobability of CV Gaussian processes and determine the extremal values of its negativity.
        For an arbitrary number of modes and sequential measurements, we derive an upper bound with a closed form [Eq. \eqref{eq:upper_bound}].
        This bound is completely determined by the covariance matrices of the measurements and the symplectic unitaries that specify the dynamics.
        We prove that the maximum of the KD negativity for $M=1$ and $N=2$ is saturated by the Gaussian states corresponding to the quadrature eigenstates [Theorem \ref{th:neg_max}].
        From this result, it follows that a maximal value of the KD negativity can be reached using only Gaussin states.
        Given the typical complexity required by the implementation of non-Gaussian states, our findings suggest that, in all the scenarios where a maximal value of the KD negativity is desirable, Gaussian states already provide enough resource to accomplish this task.
        
        %This identifies KD negativity as a resource whose maximal value is within reach of standard quantum-technological platforms.
         
        %This shows how the KD negativity is governed predominantly by the process geometry rather than by the non-Gaussian character of the probed state--- the opposite behavior to Wigner negativity, which vanishes identically on every Gaussian state \cite{HUDSON1974249} and depends on the state alone. 
        %The two quasiprobabilities therefore probe complementary notions of nonclassicality, a distinction that follows directly from the block structure of \eqref{eq:KDgauss}: the Wigner function lives on a single copy of phase space, whereas the KD distribution lives on the $N$-fold space $\Gamma_N$, sensitive to the relations between successive measurements.
        
        Several applications naturally arise from these results. 
        Since any two sequential measurements can be seen as a two-time measurement scheme, the KD negativity studied in our CV setting can be applied to study  the nonclassicality of quantum thermodynamics, in particular to work extraction and heat fluctuations (see \cite{gherardini2024quasiprobabilities,lostaglio2023kirkwood}).  
        On a different note, in Ref. \cite{arvidsson2020quantum}, it is shown how the KD distribution has a direct connection with quantum metrology, where the post-selected quantum Fisher information descreases with the KD negativity of the process.
        In this sense, our findings translate into an upper bound on the variance of a parameter estimation protocol, using Gaussian process.
        This observation suggests that, for quantum metrology, Gaussian resources are often enough to achieve quantum advantage \cite{caves1981quantum}.
        We leave some open questions for further investigation.
        On the one hand, our results could provide useful insights into the analysis of the coarse-graining property of the KD quasiprobability \cite{jia2026temporal}.
        On the other, the extension of Theorem \ref{th:neg_max} to the multi-mode domain could shed light on the link between negativity, Gaussian entanglement, and anomalous energy exchanges between modes \cite{gherardini2024quasiprobabilities}.
        Such a study could take advantage of group theoretical arguments, as found for the single-mode case. 
    
\paragraph{Acknowledgments.---}
    LB thanks E. Trombetti, S. Gherardini, K. Tanji, and W. Roga for useful discussions.
    LB acknowledges support from the European Union ERC StG, QOMUNE, 101077917.

\bibliography{biblio}
\end{document}

\label{SM}

\title{
    Supplemental Material of ``Bounding Kirkwood-Dirac negativity of Gaussian processes"
}

\author{Luca Bianchi}
\affiliation{Department of Physics and Astronomy, University of Florence, 50019, Firenze, Italy}

\author{Carlo Marconi}
\affiliation{Istituto Nazionale di Ottica del Consiglio Nazionale delle Ricerche (CNR-INO), 50125 Firenze, Italy}

\author{Riccardo Cioli}
\affiliation{
Dipartimento di Fisica e Astronomia, Università di Bologna and INFN, Sezione di Bologna, via Irnerio 46, I-40126 Bologna, Italy
}

\author{Jan Sperling}
%\email{jan.sperling@upb.de}
\affiliation{Theoretical Quantum Science, Institute for Photonic Quantum Systems (PhoQS), Paderborn University, Warburger Stra\ss{}e 100, 33098 Paderborn, Germany}  

\date{\today}

\begin{abstract}
   In this Supplemental Material, we provide detailed insights into the analytical computations performed to prove the main results of the work.
   In particular, in Sec. \ref{appendix:global_up}, we prove the global upper bound for the Kirkwood-Dirac quasiprobability of Gaussian measurements.
   In Sec. \ref{appendixA:gauss}, we derive the analytical expression for the Kirkwood-Dirac of a Gaussian input state.
   In Sec. \ref{appendixB:negativity}, we prove how the negativity for two-measurements Gaussian process is a Loewner non-increasing function of the Gaussian input state, thereby allowing for the maximization procedure to be reduced to the set of pure states.
   In Sec. \ref{appendixC:proofNeg}, we provide the proof of the theorem concerning the extremal point of the Kirkwood-Dirac negativity for single-mode, two-measurements Gaussian processes.
   Eventually, in Sec. \ref{appendix: coherent}, we discuss details the Kirkwood-Dirac of superposition of coherent states.
\end{abstract}
\maketitle

\section{Proof of Theorem 1}
    \label{appendix:global_up}
    
    Since the Kirkwood-Dirac (KD) is a linear map from operator space to scalar complex function, by exploiting the convexity of states and measurements, it is sufficient to focus on pure elements, viz. 
    \begin{equation}
        Q(r_1,\dots,r_N) 
        = 
        \mathrm{Tr}[\hat{\Pi}_N \dots \hat{\Pi}_1 \ket{\phi}\bra{\phi}]
    \end{equation}
    with $\hat{\Pi}_i = \ket{\psi_i}\bra{\psi_i}$ pure Gaussian projectors on the Gaussian states $\ket{\psi_i}$ and $\ket{\phi}$ a generic pure state. 
    Let us start from $N=2$, the generalization to arbitrary $N>2$ will then be straightforward.
    For $N=2$, the function to be bounded is 
    \begin{equation}
        \mathcal{N}
        =
        \int_{\Gamma^2} dr_1dr_2
        |\braket{\phi}{\psi_2}
        \braket{\psi_2}{\psi_1}
        \braket{\psi_1}{\phi}|.
    \end{equation}
    The integrand can be symmetrically split as 
    \begin{equation}
        \mathcal{N}
        =
        \int_{\Gamma^2} dr_1dr_2
        |\braket{\phi}{\psi_2}
        \braket{\psi_2}{\psi_1}^{\frac{1}{2}}|
        |
        \braket{\psi_2}{\psi_1}^{\frac{1}{2}}
        \braket{\psi_1}{\phi}|
    \end{equation}
    which can be upper-bounded by the Cauchy-Schwartz inequality
    \begin{equation}
        \label{eq:CS}
        \mathcal{N}
        \leq 
        \left(
            \int_{\Gamma^2} dr_1dr_2
                |\braket{\phi}{\psi_2} |^2
                |\braket{\psi_2}{\psi_1} |
        \right)^{\frac{1}{2}}
        \left(
            \int_{\Gamma^2} dr_1dr_2
                |\braket{\psi_2}{\psi_1} |
                |\braket{\psi_2}{\phi} |^2
        \right)^{\frac{1}{2}}.
    \end{equation}
    The Gaussian kernel $|\braket{\psi_2}{\psi_1}|$ is the square-root of the Gaussian envelope between the measurements, i.e.,
    \begin{equation}
        \label{def:gauss_kernel}
        |\braket{\psi_2}{\psi_1}|
        =
        \frac{
                \sqrt{2^M}
            }
            {
                \sqrt{
                        \det{
                            [\sigma_1+\sigma_2]
                        }
                    }
            }
            \exp{
                \left[
                -\frac{1}{2}(r_2-r_1)^{\T}
                (\sigma_1+\sigma_2)^{-1}
                (r_2-r_1)
                \right]
            }.
    \end{equation}
    Considering, for instance, the first term on the right-hand side (RHS) of Eq. \eqref{eq:CS}, one can shift the variable $r_1 \rightarrow \delta_{21} = r_2-r_1$ and integrate in $\delta_{21}$, leaving us with
    \begin{equation}       
            \int_{\Gamma^2} dr_1dr_2
                |\braket{\phi}{\psi_2} |^2
                |\braket{\psi_2}{\psi_1} |
        =
        (2\pi)^M 
        \sqrt[4]{
            \det[
            \sigma_1+\sigma_2
            ]}
        \int_{\Gamma} dr_2
            \braket{\phi}{\psi_2}
            \braket{\psi_2}{\phi}.
    \end{equation}
    The latter integral carries a $(2\pi)^M$ factor due to the completeness relation of the projector $\hat{\Pi}_2$. 
    Taking the square-root leads to a value of $(2\pi)^M\sqrt[8]{\det[\sigma_1+\sigma_2]}$.
    The second term on the RHS of \eqref{eq:CS} is equivalent to the first, with the only difference that $\ket{\psi_1}\rightarrow \ket{\psi_2}$ for the scalar product with the state vector $\ket{\phi}$. 
    Therefore, the universal bound on the KD negativity is given by
    \begin{equation}
        \mathcal{N}
        \leq 
        (2\pi)^{2M}\sqrt[4]{\det(\sigma_1+\sigma_2)}.
    \end{equation}

    The generalization to arbitrary $N$ is performed by splitting the negativity integrand into
    \begin{equation}
        \int_{\Gamma^N} dr 
        | \braket{\phi}{\psi_N}| 
        |\braket{\psi_N}{\psi_{N-1}} \dots \braket{\psi_1}{\phi}|^{\frac{1}{2}} 
        | \braket{\psi_N}{\psi_{N-1}} \dots \braket{\psi_1}{\phi}|^{\frac{1}{2}}|
        |\braket{\phi}{\psi_1}|
    \end{equation}
    and applying again Cauchy-Schwartz to get
    \begin{equation}
        \mathcal{N} 
        \leq
        \left(
            \int_{\Gamma^N} dr
                \braket{\phi}{\psi_N}\braket{\psi_N}{\phi}
                |
                    \braket{\psi_N}{\psi_{N-1}}
                    \dots 
                    \braket{\psi_2}{\psi_1}
                |
        \right)^{\frac{1}{2}}
        \left(
            \int_{\Gamma^N} dr
                \braket{\phi}{\psi_1}\braket{\psi_1}{\phi}
                |
                    \braket{\psi_N}{\psi_{N-1}}
                    \dots 
                    \braket{\psi_2}{\psi_1}
                |
        \right)^{\frac{1}{2}}.
    \end{equation}
    By symmetry arguments analogous to the ones previously used, one may treat only the first integral, the second yielding the same result. 
    By iteratively applying \eqref{def:gauss_kernel}, the $N-1$ Gaussian envelope reads as
    \begin{equation}
        |
            \braket{\psi_N}{\psi_{N-1}}
            \dots 
            \braket{\psi_2}{\psi_1}
        |
        = 
        \frac{
            2^{ 
                \frac{M(N-1)}{2}
            }
        }{
            \prod_{i=1}^{N-1}
                \det{[\sigma_i+\sigma_{i+1}]}
        }
        \exp{
            \left[ 
                -\frac{1}{2}
                \sum_{i=1}^{N-1}
                    (r_{i+1}-r_i)^{\T}
                    (\sigma_i+\sigma_{i+1})^{-1}
                    (r_{i+1}-r_i)
            \right]
        }.
    \end{equation}
    One can now take the integrals over the variables $r_1,\dots,r_{N-1}$ and apply the change of variables $\delta_{i+1,i} = r_{i+1}-r_{i}$ to obtain $N-1$ Gaussian integrals of the kind 
    \begin{equation}
        \int_{\Gamma} d\delta_{i+1,i} 
            e^{-\frac{1}{2}
                \delta_{i+1,i}^{\T}
                (\sigma_{i+1}+\sigma_{i})^{-1}
                \delta_{i+1,i} 
            }
        =
        \left(\sqrt{2}\pi\right)^M
        \sqrt{
            \det[ 
                \sigma_i+\sigma_{i+1}
                ]
            }.
    \end{equation}
    By collecting all the results, the global upper bound on KD negativity for $M$ modes and $N$ Gaussian measurements is
    \begin{equation}
        \mathcal{N}
        \leq 
        (2\pi)^{MN}
        \sqrt[4]{
            \prod_{i=1}^{N-1}
            \det[ 
                \sigma_i+\sigma_{i+1}    
            ]
        }.
    \end{equation}
    Finally, adding Gaussian unitaries only rotates covariance matrices in phase space by means of the automorphism 
    $
    \hat{U}^{\dagger}_{G_i}\hat{D}_r\hat{U}_{G_i} 
    = 
    \hat{D}_{G_i^{-1}r}
    $.
    Inserting Gaussian unitary evolution $\hat{U}_{G_i}$ between each measurement $\hat\Pi_i$ allows us to promote each $\sigma_i \rightarrow \sigma^{(\mathrm{G}_i)}$  with $\mathrm{G}_i = G_{i-1} \dots G_0$ and $G_1 = G_0 = \mathrm{id}_{2M}$.
    
\section{KD for bosonic Gaussian processes}
    \label{appendixA:gauss}
    
        To prove that 
        \begin{equation}
            Q(r) = \frac{2^{MN}}{\sqrt{\det[\Lambda]}}
                    e^{
                            - r^{\T}
                            \Lambda^{-1}
                            r 
                        }
        \end{equation}
        for Gaussian states, one can expand the operators in 
        $
        Q(r) = 
                    \mathrm{Tr}
                    \left[
                        \hat F(r)
                        \hat\rho
                    \right]$
        in the Weyl representation.
        In what follows, it is useful to apply the identity
        $
        \hat{U}^{\dagger}_{G_i}\hat{D}_r\hat{U}_{G_i} 
        = 
        \hat{D}_{G_i^{-1}r}$, 
        for 
        $
        G_i \in \mathrm{Sp}_{2M}(\mathbb{R}) \, , \forall \, i=1,\dots,N-1
        $.
        The KD distribution now reads
        \begin{equation}
            \begin{split}
                Q(r) 
                &=
                    \int_{\Gamma^N} 
                        dq  
                    \int_{\Gamma} dq_0    \prod_{i=0}^N
                            \left[
                                \frac{\chi_{i}(q_i)}{(2\pi)^{M}}
                            \right]
                        \mathrm{Tr}
                            \left[
                                \hat{D}_{q_N}
                                \hat{U}_{G_{N-1}}
                                \hat{D}_{q_{N-1}}
                    \hat{U}_{G_{N-2}}  
                                \dots
                     \hat{U}^{\dagger}_{G_1}
                    \hat{D}_{q_1}
                                \hat D_{q_0}
                                \hat{U}^{\dagger}_{G_1}
                                \dots
                                \hat{U}^{\dagger}_{G_{N-1}}
                    \right]\\
                &=
                    \int_{\Gamma^N} 
                        dq
                    \int_{\Gamma} dq_0
                        \prod_{i=0}^N
                            \left[
                                \frac{\chi_{i}(q_i)}{(2\pi)^{M}}
                            \right]
                        \mathrm{Tr}
                            \left[
                                \hat{D}_{\mathrm{G}^{-1}_N q_N}
                    \hat{D}_{\mathrm{G}^{-1}_{N-1}q_{N-1}}
                                     \dots
                                    \hat{D}_{\mathrm{G}^{-1}_1q_1}
                                    \hat D_{q_0}
                            \right].\\
            \end{split}
        \end{equation} 
        Since each $\mathrm{G}_i$ is symplectic, changing variables $\mathrm{G}^{-1}_i q_i \rightarrow q_i$ does not affect the phase-space volume.
        By means of the orthogonality relation between Weyl operators, the trace between displacements gives
        \begin{equation}
            \begin{split}
                \exp{
                    \left[
                        -\frac{i}{2}
                        \sum_{j>k}^{N}
                            q_j^{\T}\Omega q_i
                    \right]
                    }
                \mathrm{Tr}
                \left[
                    \hat{D}
                        \left(
                            \sum_{i=0}^Nq_i
                        \right)
                \right]
                =
                \exp{
                    \left[
                        -\frac{i}{4}\left(
                        \sum_{j>k}^{N}
                            q_j^{\T}\Omega q_k
                        -\sum_{j<k}^{N}
                            q_j^{\T}\Omega q_k
                        \right)
                    \right]
                    }
                \delta 
                    \left(
                        \sum_{i=0}^Nq_i
                    \right).
            \end{split}
        \end{equation}
        With the above multidimensional Dirac delta function, one variable can be fixed to be combination of the others, viz. $q_0 = - \sum_{i=1}^Nq_i$. 
        The KD distribution then becomes
        \begin{equation}
            \label{eq: KD_general}
                \begin{split}
                    Q(r) =
                        \int_{\Gamma^N}
                        dq 
                        \prod_{i=1}^N 
                            \left[
                                \frac{\chi_{i}(\mathrm{G}_iq_i)}{(2\pi)^M} 
                            \right]
                        \chi_0\left(
                                -\sum_{i=1}^Nq_i
                            \right)
                    \exp{
                    \left[
                        -\frac{i}{4}\left(
                        \sum_{j>k}^{N}
                            q_j^{\T}\Omega q_k
                        -\sum_{j<k}^{N}
                            q_j^{\T}\Omega q_k
                        \right)
                    \right]
                    }.
                \end{split}
        \end{equation}
        By using the expression for a Gaussian characteristic function, one can compute
        \begin{equation}
        \label{proof}
            \begin{split}
                Q(r) 
                &=
                    \int_{\Gamma^N} \frac{dq}{(2\pi)^{MN}}
                    \exp{
                        \left[ 
                            -\frac{1}{4}
                            \sum_{j.k=1}^Nq_j^{\T}\Omega^{\T}
                                \Lambda
                                \Omega q_k
                            +i\sum_{k=1}^N q_k^{\T}\Omega^{\T} 
                                \left(
                                    r^{(\mathrm{G}_k)} - r
                                \right)
                        \right]
                    }\\
                    &= \frac{2^{MN}}{\sqrt{\det[\Lambda]}}
                    \exp{
                        \left[
                            - r^{\T}\Lambda^{-1}r 
                            \right]
                        } 
            \end{split}
        \end{equation}
        where 
        $
        \Lambda 
        = 
        \sigma^{(\mathrm{G}_j)}\delta_{jk}
        + \sigma 
        -i\Omega\sum_{l=1}^N\delta_{j-k, l}
        +i\Omega\sum_{l=1}^N\delta_{k-j, l}
        $
        and 
        $
        \sigma^{(\mathrm{G}_j)} 
        =
        (\mathrm{G}\mathrm{G}^{\T})^{-\T}_j
        \sigma_j
        (\mathrm{G}\mathrm{G}^{\T})_j
        $
        and 
        $\mathrm{G}_j = G_{j-1} \dots G_0$.
        The above integral always converges as long as $v^{\T}\mathrm{Re}[\Lambda]v > 0 \; \forall \, v\neq 0$.
        Since $\mathrm{Re}[\Lambda] = \bigoplus_{i=1}^N\sigma_i + \sigma\mathbf{1}\mathbf{1}^{\T}$ ($\mathbf{1} = (1,\dots,1)^{\T}$), this gives $\sum_{i=1}^N v_i^{\T}\sigma_i v_i \geq 0$ 
        and 
        $
        \left(\sum_{i=1}^N v_i\right)^{\T}\sigma\left(\sum_{j=1}^N v_j\right) \geq 0
        $,
        implying the convergence of the integral as long as one of the covariance matrices involved has finite (squeezing) eigenvalues.
        In particular, for the maximization procedure considered in this work, the assumption
        $
        \mathrm{Tr}[\sigma_i]<\infty \; \forall \, i=1,\dots,N
        $
        is imposed.
        As a final remark, notice how the term
        $
        \exp{
            \left[ 
                -\frac{1}{4}            \sum_{j.k=1}^N
                q_j^{\T}
                \Omega^{\T}
                \Lambda
                \Omega 
                q_k
            \right]
            } 
        $
        inside the integral of Eq. \eqref{proof} can be interpreted as a generalized Gaussian characteristic function.
        In fact, a genuine, properly defined characteristic function $\chi(q)$ should satisfy the positivity condition given by Bochner's theorem \cite{Bochner}, namely
        \begin{equation}
            \chi(q_j-q_k)e^{\frac{i}{2}q^{\T}_j\Omega q_k} \succ 0 \, \forall j,k
        \end{equation}
        which fails in this case due to the phase-space area term in the off-diagonal block structure of $\Lambda$.

\section{Loewner monotonicity of Kirkwood-Dirac negativity}
    \label{appendixB:negativity}
    
        In order to simplify computations, it is useful to consider the logarithmic version of the negativity, with a proper prefactor:
        \begin{equation}
            \label{eq:def_negativity}
                \tilde{ \mathcal{N}}_g = 4\log{
                    \left[
                        \int_{\Gamma^N} dr
                          \left|
                                Q(r)
                            \right|
                    \right]
                    }.
        \end{equation}
        The form of Eq. \eqref{eq:def_negativity} does not spoil any eventual ordering induced by the negativity, as it is a composition of monotonic functions of $\tilde{\mathcal{N}}_g$.
        Performing the integral in Eq. \eqref{eq:def_negativity} gives 
        \begin{equation}
            \label{eq:intermediateN}
            \tilde{\mathcal{N}}_g 
            = 
            4MN\log{
                2\pi
            } 
            +
            2\log{
                \left[ 
                    \frac{|\det[\Lambda]|}{\det[\mathrm{Re}[\Lambda]]}
                \right]
            }. 
        \end{equation}
                
        Letting $N=2$, one can exploit the invariance of $\tilde{\mathcal{N}}_{g}$ under symplectic transformations by performing a rotation $U = (1/\sqrt{2})\begin{bmatrix} \mathrm{id}_{2M} & \mathrm{id}_{2M}\\ \mathrm{id}_{2M} & -\mathrm{id}_{2M}\end{bmatrix}$ on $\Lambda$. 
        In this way, one introduces 
        \begin{equation}
            \begin{split}
            \tilde\Lambda &= U \Lambda U^{\T}
            = \begin{bmatrix}
                    2\sigma + \alpha & \beta + i\Omega \\
                    \beta -i\Omega & \alpha
                \end{bmatrix}
            \end{split},
        \end{equation}
        with 
        $
        \alpha 
        = (1/2)
        \left(
            \sigma^{(\mathrm{G}_1)}+\sigma^{(\mathrm{G}_2)}
        \right)
        $ 
        and 
        $
        \beta 
        = 
        (1/2)
        \left(
        \sigma^{(\mathrm{G}_1)}-\sigma^{(\mathrm{G}_2)}
        \right)
        $ 
        two real-symmetric matrices.
        By further simplifying the notation with 
        $
        C 
        = 
        (1/2)
        [
        \alpha - (\beta + i\Omega)\alpha^{-1}(\beta -i\Omega) 
        ]
        $ 
        and 
        $
        D 
        =
        (1/2)
        [
        \alpha -\beta\alpha^{-1}\beta
        ]
        $, the negativity becomes
        \begin{equation}
            \tilde{\mathcal{N}}_g = 
            4MN\log{
                2\pi
            } 
            +
            2\log{
                \left[ 
                    \frac{|\det[\sigma + C]|}{\det[\sigma+D]}
                \right]
            }
        \end{equation}
        and by noticing how $C = D + Z$, $Z = X+iY$, with 
        $X = -\Omega \alpha^{-1} \Omega$ 
        and 
        $Y = \beta\alpha^{-1}\Omega - \Omega\alpha^{-1}\beta$, one finally arrives at
        \begin{equation}
            \tilde{\mathcal{N}}_g 
            = 
            4MN\log{
                2\pi
            } 
            +
            \log{
                \det
                \left[ 
                    \frac{(\sigma + D + Z)(\sigma + D + Z^{\dagger})}{(\sigma + D)^2}
                \right].
            }
        \end{equation}
        From this expression, one proves the following lemma.
        \begin{lemma}
        \label{prop1}
        The KD negativity of a two-measurement Gaussian process is a Loewner-monotonically non-increasing function of the covariance matrix of the input state $\sigma$
        \end{lemma}
        
        \begin{proof}
            Let us begin by noticing that $\alpha$ is a positive definite matrix, being $\sigma_{1,2} \succ 0$ by hypothesis. 
            As a consequence, also $\beta\alpha^{-1}\beta \succ 0$, since $\beta$ real symmetric, thereby giving $D \succ 0$ and thus $\sigma + D \succ 0 $.
            An analogous argument gives $X \succ 0$ and $Y \succ 0$.
            
            To check the monotonicity in $\sigma$, one can take $ \forall \; H \succ 0$  the directional derivative
            \begin{equation}
                \left.\frac{d}{dt}\tilde{\mathcal{N}}_{g}(\sigma +tH)\right|_{t=0} 
                =
                \lim_{t\rightarrow 0}\frac{\tilde{\mathcal{N}}_g(\sigma+tH)-\tilde{\mathcal{N}}_g(\sigma)}{t}
            \end{equation}
            and test whether it is negative. 
            In particular, 
            \begin{equation}
                \tilde{\mathcal{N}}'_g(\sigma) 
                = 
                2\mathrm{Tr}[
                    (
                    \mathrm{Re}[(\sigma+D+Z)^{-1}]
                    -
                    (\sigma + D)^{-1}
                    )H
                    ]
            \end{equation}
            is negative if the matrix inside the trace is negative definite, which happens if 
            \begin{equation}
                \mathrm{Re}[(\sigma+D+Z)^{-1}] \prec (\sigma + D)^{-1}.
            \end{equation}
            To check that the above inequality always holds, we rewrite
            \begin{equation}
                (\sigma+D+X+iY)^{-1} 
                = 
                (\sigma+D+X)^{-\frac{1}{2}}
                (
                \mathrm{id}_{2M} + i\tilde Y 
                )^{-1}
                (\sigma+D+X)^{-\frac{1}{2}}
            \end{equation}
            with $
            \tilde Y 
            = 
            (\sigma+D+X)^{-\frac{1}{2}}Y(\sigma+D+X)^{-\frac{1}{2}}
            $, which is positive definite by construction.
            Calling $\lambda_k >0 \; \forall k \in \{1,\dots,2M\}$ the eigenvalues of $\tilde Y$ one can prove the following inequality
            \begin{equation}
                \begin{split}
                    (\sigma+D+X)^{-\frac{1}{2}}
                    \mathrm{Re}
                    \left[
                        U^{\dagger}\mathrm{diag}
                            \left(
                                \frac{1}{1+\lambda_{1}^2},\dots,\frac{1}{1+\lambda_{2M}^2}
                            \right)
                        U
                    \right]
                    (\sigma+D+X)^{-\frac{1}{2}}
                    \prec 
                    (\sigma+D)^{-1},
                \end{split}
            \end{equation}
            thus concluding the proof.
        \end{proof}
        
        As a consequence of Lemma \ref{prop1}, the global maximum of $\mathcal{N}_g$ lies on the border of the set of covariance matrices for Gaussian states $\sigma$;
        that is, for the global maximum, it is sufficient to explore pure states since $\sigma_{\mathrm{pure}} \prec \sigma_{\mathrm{mixed}}$ \cite{serafini2023quantum}. (Note that the same argument does not hold for the global minimum). 
        The purity condition reads $\det[\sigma] = 1$, which implies $\nu_j = 1, \forall \ j=1,\dots,M$ and reduces the independent components of $\sigma$ to two real parameters.

\section{Proof of Theorem 2}
\label{appendixC:proofNeg}

    Let us assume that the input state is a Gaussian state parametrized by the covariance matrix $\sigma$.
    The KD negativity for a Gaussian state in a Gaussian process, which here reads as
    \begin{equation}
        \label{def:negativity}
        \mathcal{N}_g(\sigma) 
        = 
        (2\pi)^2
        \sqrt{
            \frac{
                |\det{[\Lambda]}|
            }
            {\det{[\mathrm{Re}[\Lambda]]}}
        },
    \end{equation}
    is completely characterized by matrices of the form $N_{0} + \sigma\mathbf{1}\mathbf{1}^{\T}$, with $\mathbf{1} = (1,\dots,1)^{\T}$ is an $N$-dimensional vector and where the matrix $N_0$ is given by 
    $
    N_0 
    = 
    \left[\begin{smallmatrix} 
        \sigma_1 & -i\Omega \\
        i\Omega & \sigma_2
    \end{smallmatrix}\right]
    $ for the matrix $\Lambda$,
    while for $\mathrm{Re}[\Lambda]$ one has
    $
    N_{0}
    =
    \left[\begin{smallmatrix}
        \sigma_1 & 0 \\
        0 & \sigma_2
    \end{smallmatrix}\right]
    $.
    For the sake of notation, since matrices $G_j$ are just phase-space rotations, the superscript on $\sigma^{(\mathrm{G}_i)}$ is omitted.
    
    Before proceeding with the calculations, it is useful to recall some basic properties of the space of single-mode covariance matrices.
    A single-mode quantum covariance $\sigma$ is a $2\times2$ real-symmetric and positive definite matrix, satisfying the condition $\det[\sigma] \geq 1$: the Heisenberg principle in phase space.
    Parametrizing 
    \begin{equation}
        \label{def:cov_mink}
        \sigma 
        = 
        \begin{pmatrix}
            t+x & y \\
            y & t-x
        \end{pmatrix}
    \end{equation}
    with $t,x >0$ and calling $\det[\sigma]=\nu^2$, it is straightforward to check that
    \begin{equation}
        \label{def:minkowski}
            \nu^2 = t^2-x^2-y^2.
    \end{equation}
    That is, the parameters defining the covariance matrices live in a Minkowski space with $\eta_{ij} = \mathrm{diag}(+,-,-)$, thereby defining hyperboloids, labeled by $\nu$, over which $p^i = (t,x,y)$ defines coordinates. 
    The positivity condition implies considering only the upper sheet of any hyperboloid.
    The Williamson decomposition \cite{houde2024matrix} states that an $M$-dimensional covariance matrix $\sigma$ can be decomposed in its normal modes by means of a symplectic matrix $S$, viz. $S\bigoplus_{j=1}^M \nu_j \mathrm{id}_2 S^{\T}$, with $\nu_j$ the symplectic eigenvalues of $\sigma$, which is $\nu = 1$ for pure states and $\nu > 1$ for mixed ones. 
    Furthermore, the Bloch-Messiah decomposition \cite{simon1999congruences,braunstein2005squeezing} gives a recipe to decompose a matrix $S \in \mathrm{Sp}_{2M}(\mathbb{R})$ by means of two rotations $L, R$ and a block-diagonal matrix of the form $\Sigma = \bigoplus_{j=1}^{M} \mathrm{diag}(e^{r_j}, e^{-r_j})$, with $r_j$ denoting the squeezing parameters.
    If one uses the Williamson decomposition combined with the Bloch-Messiah reduction formula, single-mode covariance matrices can be recast as
    \begin{equation}
        \label{eq:BM}
            \sigma 
            =
            \nu^2
            \begin{pmatrix}
                \zeta\cos{\phi}^2 +\zeta^{-1}\sin{\phi}^2  & \frac{1}{2}\sin{(-2\phi)}(\zeta-\zeta^{-1}) \\
                \frac{1}{2}\sin{(-2\phi)}(\zeta-\zeta^{-1}) & \zeta^{-1}\cos{\phi}^2 +\zeta\sin{\phi}^2  
            \end{pmatrix}\,,
    \end{equation}
    with $\zeta = e^{2r}$.
    From Eq. \eqref{eq:BM}, one can also define hyperbolic coordinates in the following way:
    \begin{equation}
        \label{def:hyper}
            \begin{cases}
                t = \cosh{\beta}\\
                x = \cos{\psi}\sinh{\beta}\\
                y = \sin{\psi}\sinh{\beta}\\
            \end{cases},
    \end{equation}
    where $\beta = \ln{\zeta}$ and $\psi = -2\phi$.
    The above parametrization makes the connection between covariance matrices and events in a $(2+1)$-Minkowski spacetime explicit, with squeezing and rotations playing the role of Lorentz transformations, $\nu$ the role of a mass and \eqref{def:minkowski} a mass-shell condition. 
    In this analogy, boosting at the speed of light is equivalent to having infinite squeezing.
    Furthermore, as proved in Lemma \ref{prop1}, the KD negativity is a Loewner-monotonically non-increasing monotonic function in the space of covariance matrices;
    therefore, it is enough to restrict to pure states for the global maximum.
    For the global minimum, instead, the ordering argument cannot be used as one can have an arbitrary amount of thermal noise $\nu$.
    The minimum found in the following proof is thus attained only along the hyperboloid of pure states.
    Finally, one recalls the following facts about covariance matrices, which will be useful later:
    \begin{itemize}
        \item $\frac{1}{2}\mathrm{Tr}[\mathrm{Adj}[\sigma_1]\sigma_2] = \eta_{ij}p^i_{\sigma_1}p^j_{\sigma_2}$, with $\mathrm{Adj}[\sigma]$ the adjugate of the matrix $\sigma$. 
        This relation is analogous as the one for the distance of two $2\times2$ matrices with respect to the Hilbert-Schmidt scalar product.
        \item $\sigma\Omega = \Omega\sigma^{-1}$ and the related $\Omega\sigma = \sigma^{-1}\Omega$, both coming from the definition of symplectic matrices.
        \item Any $2\times 2$ real matrix $M$ can be decomposed as $M = \alpha \ \mathrm{id} + \beta \sigma_x +\gamma \sigma_z+\delta \Omega$, with real coefficients and $\Omega$ playing the role of $\sigma_y$ for real matrices.
        \item $\det{[M_1+M_2]} = \eta_{ij}p_{M_1}^ip_{M_1}^j+\eta_{ij}p_{M_2}^ip_{M_2}^j+\eta_{ij}p_{M_1}^ip_{M_2}^j$.
        This comes from the very first observation that $\det[\sigma] = t^2-x^2-y^2 = \eta_{ij}p^{i}p^{j}$ together with decomposing the determinant of the sum using formulae reported in the Appendix of \cite{bohmann2016gaussian}.
    \end{itemize}

    Let us start by rewriting
    \begin{equation}
        \begin{split}
            \det[\Lambda] 
            =& 
            \det[N_0 + \sigma\mathbf{1}\mathbf{1}^{\T}]\\
            =&
            \det[N_0]
            \det[
                \mathrm{id}
                +
                \mathbf{1}^{\T}
                N_0^{-1}
                \mathbf{1}
                \sigma
            ]    
            =\\
            =&
            \det[N_0]
            \left(
                1 
                +
                \mathrm{Tr}[
                    \mathbf{1}^{\T}
                    N_0^{-1}
                    \mathbf{1}
                    \sigma
                ]
                +
                \det[
                    \mathbf{1}^{\T}
                    N_0^{-1}
                    \mathbf{1}
                    \sigma
                ]
            \right).
        \end{split}
    \end{equation}
    Therein, from the second to the third line, the matrix determinant lemma \cite{harville1998matrix} was employed, while another well-known identity for $2\times2$ matrices \cite{bohmann2016gaussian} was used for the last line. 

    To work out the action of $N_0^{-1}\mathbf{1}$, one can set up the following system
    \begin{equation}
        N_0\mathbf{w} = \mathbf{1}  
    \end{equation}
    for $\mathbf{w} = (u,v)^{\T}$ and $u,v$ two-dimensional matrices. 
    This amounts to solving a linear system for two-dimensional variables:
    \begin{equation}
        \begin{cases}
            \sigma_1 u - i\Omega v = \mathrm{id}, \\
            i\Omega u + \sigma_2 v = \mathrm{id},
        \end{cases}
    \end{equation}
    % $v = -i\Omega(\mathrm{id}-\sigma_1 u)$,
    whose solutions are $u = (\sigma_1+\sigma_2)^{-1}(\mathrm{id}+i\Omega\sigma_2^{-1})$ and $v = (\sigma_1+\sigma_2)^{-1}(\mathrm{id}-i\Omega\sigma_1^{-1})$.
    %\begin{equation}
     %   \begin{split}
      %      &i\Omega u - i\sigma_2\Omega + i\sigma_2\Omega\sigma_1 u = \mathrm{id} \\
          %  &i\Omega(\mathrm{id}+\sigma_2^{-}\sigma_1)u = \mathrm{id}+i\sigma_2\Omega \\
        %    &i\Omega\sigma_2^{-}(\sigma_1+\sigma_2)u = \mathrm{id}+i\sigma_2\Omega \\
            %&i\sigma_2\Omega(\sigma_1+\sigma_2)u = \mathrm{id}+i\sigma_2\Omega\\
         %   &u = (\sigma_1+\sigma_2)^{-}(\mathrm{id}+i\Omega\sigma_2^{-})
       % \end{split}
    %\end{equation}
    %while from $u = i\Omega(\mathrm{id}-\sigma_2v)$ one gets to  with analogous computations.
    By defining $\Sigma = \sigma_1+\sigma_2$ and $\Delta = \sigma_2-\sigma_1$, one then has $u+v = \Sigma^{-1}(2\mathrm{id}+i\Delta\Omega)$, which leads to
    \begin{equation}
        \begin{split}
        \det[\Lambda]
        =&
        \det[N_0]
        (
            1
            +
            \mathrm{Tr}[
                \sigma 
                [
                    \Sigma^{-1}
                    (
                        2\mathrm{id}
                        +
                        i\Delta\Omega
                    )
                ]
            ]
            +\det[\Sigma^{-1}]
            \det[ 
                  2\mathrm{id}
                    +
                    i\Delta\Omega
            ],
        \end{split}
    \end{equation}
    where $\det[\sigma]=1$ was used.
    
    Now we firstly address the determinant term.
    By employing again well-known relations between determinants of $2\times 2$ matrices \cite{bohmann2016gaussian}, it is possible to write
    \begin{equation}
        \det[
            2\mathrm{id}
            +
            i\Delta\Omega
            ]
            =
            \det[ 
            2\mathrm{id}
            ]
            +
            \det[ 
            \Delta\Omega
            ]
            +
            2i 
            \mathrm{Tr}
            [
            \Omega \Delta
            ].
    \end{equation}
    The last term vanishes as it is the trace of the product of a symmetric matrix with an antisymmetric one, while the second term gives $-\det[\Delta]$, yielding a total determinant of $4-\det[\Delta]$.
    Furthermore, one can rewrite
    $
    \det[\Delta] 
    = 
    \eta_{ij}p^{i}_{\Delta}p^{j}_{\Delta} 
    = 
    2(1-\eta_{ij}p^{i}_{1}p^{j}_{2})
    $ 
    and 
    $
    \det[\Sigma] 
    = 
    \eta_{ij}p^{i}_{\Sigma}p^{j}_{\Sigma}
    =
    2(1+\eta_{ij}p^{i}_{1}p^{j}_{2})
    $
    In particular, it turns out that $4-\det[\Delta] = \det[\Sigma]$, thereby leaving
    $
    \det[\Sigma^{-1}]\det[ 
                  2\mathrm{id}
                    +
                    i\Delta\Omega
            ] = 1.
    $
    
    To address the trace, which gives the real and imaginary parts of $\det[\Lambda]$, one can show that 
    $
        \det[N_0]
        =
        \det[\Sigma].
    $
    To prove this, the purity of covariance matrices is used together with $\mathrm{Tr}[\sigma_1^{-1}\sigma_2] = 2\eta_{ij}p_1^{i}p_2^{j}$ which leads to an overall $2(1+\eta_{ij}p_1^{i}p_2^{j}) = \det[\Sigma]$ term.
    Therefore, the first two pieces of the trace term give
    \begin{equation}
        \det[\Sigma] + \mathrm{Tr}[\mathrm{Adj}[\Sigma]\sigma] 
        =
        2(\eta_{ij}p_1^{i}p_2^{j} + 1) 
        +
        2(\eta_{ij}p_1^{i}p_1^{j} + \eta_{ij}p_2^{i}p_2^{j}).
    \end{equation}
    The last term in the trace can be rewritten as
    \begin{equation}
        \mathrm{Tr}[
            \sigma\mathrm{Adj}[\Sigma]\Delta\Omega
        ].
    \end{equation}
    By decomposing the matrix $\sigma\mathrm{Adj}[\Sigma]\Delta$ in the Pauli basis, only the purely anti-symmetric term survives. 
    After further algebra, one readily recognizes that  
    \begin{equation}
        \label{eq:determinant}
        \mathrm{Tr}[
            \sigma\mathrm{Adj}[\Sigma]\Delta\Omega
        ]
        = 
        \det{ 
            \begin{pmatrix}
                t_{\sigma} & x_{\sigma} & y_{\sigma} \\
                t_{\Sigma} & x_{\Sigma} & y_{\Sigma} \\
                t_{\Delta} & x_{\Delta} & y_{\Delta}
            \end{pmatrix}
            }.
    \end{equation}
    Now, using the short-hand notation $g_{\mu\nu} = \eta_{ij}p_{\mu}^{i}p_{\nu}^{j}$, calling $U=1+g_{11}+g_{22}+g_{12}$ and setting $V = \det\left(\begin{smallmatrix}
        t_{\sigma} & x_{\sigma} & y_{\sigma} \\
        t_1 & x_1 & y_1 \\
        t_2 & x_2 & y_2
    \end{smallmatrix}\right)
    $ 
    one arrives at
    \begin{equation}
        |\det{[\Lambda]}| 
        =
        2\sqrt{
            U(\sigma)^2+V(\sigma)^2
        }.
    \end{equation}

    Analogous computations can be performed for $\det[\mathrm{Re}[\Lambda]]$, which is simpler than $\det[\Lambda]$ since $N_0$ is now block-diagonal,
    \begin{equation}
        \begin{split}
            |\det{[\mathrm{Re}[\Lambda}]]| 
            =&
            1
            +
            \mathrm{Tr}[
                \mathrm{Adj}[\Sigma]
                \sigma 
            ]
            +
            \det{[\Sigma]} \\
            =& 1 + 2(g_1 + g_2) + 2 + 2g_{12}\\
            =& 2U(\sigma) + 1.
        \end{split}
    \end{equation}

    The negativity now is on a fully manifest Lorentz invariant form, being $U$ and $V$ Lorentz scalars,
    \begin{equation}
        \mathcal{N}_g(\sigma) 
        =
        \sqrt{2}(2\pi)^2
        \sqrt{
         \frac{
                \sqrt{
                    U(\sigma)^2+V(\sigma)^2
            }
        }
        {
            2U(\sigma) + 1
        }
        }.
    \end{equation}
    The advantage of Lorentz invariance is that one can evaluate scalars in any reference frame, which amounts to apply symplectic transformations to the negativity.
    One can pursue the analogy with special relativity and apply a Lorentz transformation to land in the reference frame of the center of mass (CM) between $\sigma_1,\sigma_2$, where the terms in $p_1,p_2$ are greatly simplified. 

    To build the CM  frame, let us consider a triad $\hat{e}_t, \hat{e}_x, \hat{e}_y$.
    One vector, say $\hat{e}_t$, should be time-like $\eta_{ij}\hat{e}^i_{t}\hat{e}^j_{t} = 1$, while the other two $\hat{e}_x,\hat{e}_y$ should be space-like (Minkowski norm equal to $-1$).
    In the CM frame, the time component of the CM should be the sum of the energies $t_1,t_2$, while the spatial components of $p_1$ and $p_2$ should have equal modulus and opposite direction.
    Therefore, only $p_1,p_2$ are needed to characterize the basis $\hat{e}$.
    Starting from $\hat{e}_t \varpropto p_1+p_2$, one imposes the time-like condition to normalize this vector, obtaining
    \begin{equation}
        \hat{e}_t 
        = 
        \frac{
            p_1+p_2
        }{
            2k
        },
    \end{equation}
    where $k = \sqrt{\frac{1+g_{12}}{2}}$.
    The second vector can be taken as a space-like, orthogonal counterpart of $\hat{e}_t$, i.e., $\hat{e}_x \varpropto p_1-p_2$. 
    Again, the normalization yields
    \begin{equation}
        \hat{e}_x 
        = 
        \frac{
            p_1-p_2
        }
        {
        2\sqrt{k^2-1}
        }.
    \end{equation}
    The third vector should also be spacelike, and orthogonal to $\hat{e}_x$. 
    A natural assumption is to impose $\hat{e}_y \varpropto \hat{e}_t\, \wedge \, \hat{e}_x $, and, after algebraic manipulations and normalization, it gives
    \begin{equation}
        \hat{e}_y 
        = 
        \frac{p_1 \wedge p_2}
        {
        \sqrt{
            1-g_{12}^2
        }
        }
    \end{equation}
    One can correctly recover the components of $p_1,p_2$ in this frame 
    \begin{equation}
        \begin{cases}
            p_1 = (k, -\sqrt{k^2-1},0), \\
            p_2 = (k, \sqrt{k^2-1},0),
        \end{cases}
    \end{equation}
    which also shows how $k = \cosh{\beta_{\mathrm{rel}}}$ is related to the rapidity of the boost, $\beta_{\mathrm{rel}}$ being the relative velocity between $p_1,p_2$.
    The value of $\beta_{\mathrm{rel}}$ is not needed since one can notice $k = \cosh{\beta_{\mathrm{rel}}} = t_i^{\mathrm{CM}} = \frac{1}{2}\sqrt{\det{\Sigma}}$:
    in the CM frame, the time-component of the Lorentz vector of one of the two particles is half of the square root of the total four-momentum norm, i.e., $\frac{1}{2}\sqrt{\eta_{ij}p_{\sigma_1+\sigma_2}^{i}p_{\sigma_1+\sigma_2}^{j}}$, which is exactly $\frac{1}{2}\sqrt{\det{\Sigma}}$.
    For the input state, one gives generic coordinates with respect to which the optima should be found 
    \begin{equation}
        p = (\cosh{\beta},\cos\psi\sinh{\beta},\sin\psi\sinh{\beta}).
    \end{equation}
    The scalar functions $U(\sigma),V(\sigma)$ are particularly easy in the CM frame,
    \begin{equation}
        \begin{cases}
            U = 2k(k+\cosh{\beta}),\\
            V = -2k\sqrt{k^2-1}\sinh{\beta}\sin{\psi} .
        \end{cases}
    \end{equation}
    Now, $U^2$ does not depend on the angle $\psi$, while $V^2$ has a simple $\sin^2\psi$ dependence. 
    Therefore, the optimization with respect to $\psi$ is straightforward:
    the maximum will be achieved for $\psi = \frac{\pi}{2}$, the minimum for $\psi = 0$.

    By recalling that $t = \cosh{\beta}$, for $\psi = \frac{\pi}{2}$, the numerator of the negativity is 
    \begin{equation}
        \sqrt{4k^2(k+t)^2+(t^2-1)4k^2(k^2-1)},
    \end{equation}
    which leaves
    \begin{equation}
        \mathcal{N}_g 
        = 
        \sqrt{2}(2\pi)^2
        \sqrt{
            \frac{
                2k^2 t + 2k
            }{
            4k^2 + 4kt + 1
        }
        }.
    \end{equation}
    Deriving in $t$ (which is equivalent to derive in $\beta$) gives $N'_g \propto 4k^2(4k^2-3)$, which is always satisfied since $\cosh{\beta_{\mathrm{rel}}}\geq 1$.
    Therefore, the negativity always increases in the direction of the optimal angle, actually approaching the supremum at $t\rightarrow \infty$.
    In particular, the value fixed by the limit amounts to 
    \begin{equation}
        \begin{split}
        \mathcal{N}_g^{\mathrm{max}} 
        &= 
        (2\pi)^2\sqrt{2k}\\
        &=
        \sqrt{2}(2\pi)^2
        \left(
            \frac{1+g_{12}}{2}
        \right)^{1/4}\\
        &=
        (2\pi)^2
        \sqrt[4]{\det{[\Sigma]}}.
    \end{split}
    \end{equation}

    Analogously, the minimum is achieved for $\psi = 0$ and, after noticing that the derivative is always positive, attained for $t=1$, corresponding to $\beta = 0$.
    This leads to
    \begin{equation}
        \mathcal{N}_{g}^{\mathrm{min}} 
        = 
        (2\pi)^2
        \sqrt{
            \frac{
                4k(k+1)
            }
            {
                4k(
                k + 1
                ) + 1
            }
        },
    \end{equation}
    which is a non-trivial function of $\sigma_1,\sigma_2$.
    Substituting the value of $k$ finally gives
    \begin{equation}
        \mathcal{N}_g^{\mathrm{min}}
        =
        (2\pi)^2
        \sqrt{
            \frac{
                \sqrt{
                    \det{
                        [\Sigma]
                    }
                }(
                    \sqrt{
                    \det{
                        [\Sigma]
                    }
                } + 2
                )
            }{
                \sqrt{
                    \det{
                        [\Sigma]
                    }
                }(
                    \sqrt{
                    \det{
                        [\Sigma]
                    }
                } + 2
                )
                + 1
            }
        }.
    \end{equation}

    To obtain the location of such optima, one should boost back the values of $t,\psi$ into the lab frame. 
    Expanding $p$ in the CM frame amounts to computing simple scalar products between Lorentz vectors,
    \begin{equation}
        \begin{split}
            p^{\mathrm{CM}}
            &=
            \frac{1}{2k}\eta_{ij}(p_1+p_2)^ip^j\hat{e}_t
            +
            \frac{1}{{2\sqrt{k^2-1}}}\eta_{ij}(p_1-p_2)^ip^j\hat{e}_x
            +
            \frac{1}{{\sqrt{1-g_{12}^2}}}\eta_{ij}(p_1\wedge p_2)^i p^j\hat{e}_y\\
            &=
            \frac{g_1+g_2}{2k}\hat{e}_t
            +
            \frac{g_1-g_2}{2\sqrt{k^2-1}}\hat{e}_x
            -
            \frac{V}{\sqrt{1-g_{12}^2}}\hat{e}_y
        \end{split}.
    \end{equation}
    
    For the minimum, impose $\beta=0$ and $\psi = \pi/2$ in the vector $p$, compute the invariants $g_1,g_2,V$, and find the coefficients in the above expansion. 
    It turns out that the minimum in the lab frame is attained for 
    \begin{equation}
        p^{\mathrm{min}} 
        = 
        \frac{p_1+p_2}{2k}\hat{e}_t 
        =
        \left(
        \frac{t_1+t_2}{2k}, \frac{x_1+x_2}{2k}, \frac{y_1+y_2}{2k}
        \right),
    \end{equation}
    thereby giving 
    \begin{equation}
        \begin{cases}
        \beta^{\mathrm{min}} 
        = 
        \frac{
            \cosh{\beta_1} + \cosh{\beta_2}
        }{
            2\cosh{\beta_{\mathrm{rel}}}
        },
        \\
        \psi^{\mathrm{min}} 
        = 
        \arctan{
            \left[ 
                \frac{
                    \sinh{\beta_1}\sin{\psi_1}+\sinh{\beta_2}\sin{\psi_2}
                }
                {
                \sinh{\beta_1}\cos{\psi_1}+\sinh{\beta_2}\cos{\psi_2}
                }
            \right]
        }.
        \end{cases}
    \end{equation}
    For the supremum, impose $\beta \rightarrow \infty$ and $\psi = 0$, evaluate the invariants, and find the lab components
    \begin{equation}
        p^{\mathrm{max}} 
        = 
        \frac{\cosh{\beta}}{2k}(p_1+p_2)
        \hat{e}_t 
        +
        \frac{k\sinh{\rho}}{\sqrt{1-g_{12}^2}}(p_1\wedge p_2)
        \hat{e}_y,
    \end{equation}
    giving 
    \begin{equation}
        \begin{cases}
            \beta^{\mathrm{max}} \rightarrow \infty, \\

            \psi^{\mathrm{max}} 
            = 
            \arctan{
                \left[ 
                    \frac{
                        \sinh{\beta_1}\sin{\psi_1} 
                        + 
                        \sinh{\beta_2}\sin{\psi_2}
                        +
                        \frac{2k^2}
                        {
                        \sqrt{1-g_{12}^2}
                        }
                        (
                        \cosh{\beta_2}\sinh{\beta_1}\cos{\psi_1}
                        -
                        \cosh{\beta_1}\sinh{\beta_2}\cos{\psi_2}
                        
                        )
                    }
                    {

                    \sinh{\beta_1}\sin{\psi_1} 
                        +
                    \sinh{\beta_2}\sin{\psi_2}
                    +
                        \frac{2k^2}
                        {
                        \sqrt{1-g_{12}^2}}
                        (
                            \cosh{\beta_1}
                            \sinh{\beta_2}
                            \sin{\psi_2}
                            -
                            \cosh{\beta_2}
                            \sinh{\beta_1}
                            \sin{\psi_1}
                        )
                    }
                \right]
            }.
        \end{cases}
    \end{equation}
    To come back to the original Bloch-Messiah coefficients, one simply substitutes back $\psi_i = -2\phi_i$ and $\beta_i = \ln\zeta_i$.
    
\section{Superposition of coherent states}
        \label{appendix: coherent}

            The most general $M$-mode superposition of coherent states can be written as
                \begin{equation}
                 \label{def:gen_cat}
                    \ket{\psi} 
                    = 
                    \sum_{i=1}^{s}
                    c_i\ket{\alpha_i},
                \end{equation}
                where each $\alpha_i$ is an $M$-dimensional vector.
                Because coherent states span the entire Hilbert space, as $s\rightarrow \infty$, this superposition can approximate any state arbitrarily well.
            The density matrix has the form
            \begin{equation}
            \label{def:rho_gen}
                \rho_{\psi} 
                =
                \sum_{i=1}^{s}|c_i|^2 \ket{\alpha_i}\bra{\alpha_i} +
                \sum_{i\neq j}^{r} c_i^* c_j \ket{\alpha_i}\bra{\alpha_j}
            \end{equation}
            In the following, ``$||\cdot||$'' stands for the modulus of real vectors, reserving ``$|\cdot|$'' to the modulus of a complex number.
            
            The characteristic function of the state in Eq. \eqref{def:rho_gen} can be computed as
            \begin{equation}
                \label{eq:general_chat}
                \begin{split}
                    \chi_{\psi}(\beta) 
                    &= 
                    \mathrm{Tr}[\hat{D}_{\beta}\hat\rho_{\psi}]\\
                    &=
                    \sum_{i=1}^{s}
                        |c_i|^2 \bra{\alpha_i}\hat{D}_{\beta}\ket{\alpha_i} +
                    \sum_{i\neq j}^{r}
                        c_i^*c_j \bra{\alpha_i}\hat{D}_{\beta}\ket{\alpha_j}\\
                    &=
                    \sum_{i=1}^{s}
                        |c_i|^2
                        e^{
                            \frac{1}{2}(\beta\alpha_i^*-\alpha_i\beta^*)
                        }
                        \braket{\alpha_i}{\alpha_i+\beta}
                    +\sum_{i\neq j}^{s}
                        c_i^*c_j 
                        e^{
                            \frac{1}{2}(\beta\alpha_j^*-\alpha_j\beta^*)
                        }
                        \braket{\alpha_i}{\alpha_j+\beta}\\
                    &=
                    \sum_{i=1}^{s}
                        |c_i|^2
                        e^{
                            \frac{|\beta|^2}{2}
                        }
                        e^{
                            2i\mathrm{Im}[\beta\alpha_i^*]
                        }\\
                    &\quad+
                    \sum_{i\neq j}^{s}
                        c_i^*c_j 
                        e^{
                            \frac{1}{2}[\beta
                            (\alpha_j^*-\alpha_i^*)-(\alpha_j-\alpha_i)\beta^*]
                        }
                        \exp{
                            \left[
                            -\frac{|\beta+\alpha_j-\alpha_i|^2}{2} +
                        \frac{1}{2}
                            (\beta + \alpha_j)\alpha_i^* - 
                            (\beta^*+\alpha_j^*)\alpha_i
                        \right]
                        }\\
                     &=
                     e^{
                        -\frac{|\beta|^2}{2}
                     }
                     \left(
                        \sum_{i=1}^{s}
                            |c_i|^2
                            e^{
                                2 i \mathrm{Im}(\beta\alpha_i^*)
                            }+   
                        \sum_{i\neq j}^{s}
                            c_i^*c_j
                            e^{
                            -\frac{|\alpha_i|^2}{2} - \frac{|\alpha_j|^2}{2} +\alpha_j^*\alpha_i
                            }
                            e^{
                                \beta\alpha_i^* -\beta^*\alpha_j
                            }
                     \right),
                \end{split}
            \end{equation}
            where the relations $\hat{D}_{\beta}\hat{D}_{\alpha} = e^{i\mathrm{Im}(\beta\alpha^*)} \hat{D}_{\alpha+\beta}$ and $\braket{\alpha}{\beta} = e^{-\frac{|\beta-\alpha|^2}{2}}e^{i\mathrm{Im}(\beta\alpha^*)}$ are employed.
            Finally, $\chi_{\psi}(\beta)$ can be put in the form of Eq. \eqref{eq:ch_cat} by simply calling $\tilde\alpha = (\mathrm{Re}(\alpha), \mathrm{Im}(\alpha))$ and switching to quadratures by means of $\beta = (1/\sqrt{2})(x+ip)$.
            \begin{equation}
                \label{eq:ch_cat}
                \begin{split}
                    \chi_{\psi}(q) 
                    =& 
                    e^{-\frac{||q||^2}{4}}
                        \sum_{i=1}^{s}
                            |c_i|^2
                            e^{
                                i\sqrt{2}
                            q^{\T}\Omega^{\T}\tilde\alpha_i
                            }\\
                    &+e^{-\frac{||q||^2}{4}}\sum_{i\neq j}^{s}
                            c_i^* c_j
                            \exp{
                            \left\{
                                -\frac{||\tilde\alpha_i||^2}{2}
                                -\frac{||\tilde\alpha_j||^2}{2}
                                +\tilde\alpha_i^{\T}\tilde\alpha_j + i\tilde\alpha_i^{\T}\Omega\tilde\alpha_j
                            +
                                \sqrt{2}i
                                q^{\T}
                                \Omega^{\T} 
                                [
                            (\tilde\alpha_i+\tilde\alpha_j)
                                    -
                                    i\Omega(\tilde\alpha_i-\tilde\alpha_j)
                                ]
                            \right\}
                            },
                \end{split}
            \end{equation}
            with now the real $2M$-dimensional vector $\tilde\alpha_i = (\mathrm{Re}(\alpha_i)_1, \mathrm{Im}(\alpha_i)_1, \dots, \mathrm{Re}(\alpha_i)_M, \mathrm{Im}(\alpha_i)_M)$.
            The KD distribution for Eq. \eqref{def:gen_cat} can be readily rewritten by substituting Eq.\eqref{eq:ch_cat} into Eq. \eqref{eq: KD_general} and repeating the computations for a superposition of Gaussian integrals.

                The KD distribution for the state $\hat\rho_{\psi}$ can be computed and reads
                \begin{equation}
                    \label{eq:KD_general_cat}
                    \begin{split}
                        Q_{\psi}(r)
                        ={}& \frac{2^{MN}}{\sqrt{\det[\Lambda_{\psi}]}}
                            \Biggl\{
                            \sum_{i=1}^{s} |c_i|^2
                            \exp\!\left[
                                -\left(r^{(\mathrm{G})} - \sqrt{2}\,\bar\alpha_i\right)^{\T}
                                \Lambda_{\psi}^{-1}
                                \left(r^{(\mathrm{G})} - \sqrt{2}\,\bar\alpha_i\right)
                            \right] \\
                        &+ \sum_{i\neq j}^{s} c_i^{*} c_j
                            \exp\!\left[
                                -\frac{|\tilde\alpha_i|^2}{2}
                                -\frac{|\tilde\alpha_j|^2}{2}
                                +\tilde\alpha_i^{\T}\tilde\alpha_j
                                +i\,\tilde\alpha_i^{\T}\Omega\,\tilde\alpha_j
                            \right] \\
                        &\quad\times \exp\!\biggl[
                                -\left(
                                    r^{(\mathrm{G})}
                                    - \sqrt{2}\,(\bar\alpha_i+\bar\alpha_j)
                                    + i\sqrt{2}\,\Omega(\bar\alpha_i-\bar\alpha_j)
                                \right)^{\T} \\
                        &\qquad\qquad\times \Lambda_{\psi}^{-1}
                                \left(
                                    r^{(\mathrm{G})}
                                    - \sqrt{2}\,(\bar\alpha_i+\bar\alpha_j)
                                    + i\sqrt{2}\,\Omega(\bar\alpha_i-\bar\alpha_j)
                                \right)
                            \biggr]
                            \Biggr\}
                    \end{split},
                \end{equation}
                with 
                $
                \Lambda_{\psi} = 
                \begin{bmatrix}
                    \mathrm{id}+\sigma_1 & \mathrm{id} -i\Omega & \dots & \mathrm{id}-i\Omega \\
                    \mathrm{id}+i\Omega & \mathrm{id}+\sigma_2 & \dots & \mathrm{id}-i\Omega \\
                    \vdots & \vdots & \ddots & \vdots\\
                    \mathrm{id}+i\Omega & \mathrm{id}+i\Omega & \dots & \mathrm{id} + \sigma_{N}
                \end{bmatrix}
                $,
                and coherent amplitudes are now encoded in the vector
                $
                \tilde\alpha = (\mathrm{Re}[\alpha], \mathrm{Im}[\alpha])
                $,
                with
                $\bar\alpha_i = (\tilde\alpha_i, \dots, \tilde\alpha_i)$. 
\bibliography{biblio}